\pdfoutput=1
\documentclass[conference,10pt]{IEEEtran}
\IEEEoverridecommandlockouts
\usepackage{cite}
\usepackage{amsmath,amssymb,amsfonts}
\usepackage{algorithmic}
\usepackage{graphicx}
\usepackage{float}
\usepackage{subfigure}
\usepackage{textcomp}
\usepackage{xcolor}
\def\BibTeX{{\rm B\kern-.05em{\sc i\kern-.025em b}\kern-.08em
    T\kern-.1667em\lower.7ex\hbox{E}\kern-.125emX}}
\usepackage{algorithm}
\usepackage{amsthm}
\usepackage{subfigure}
\usepackage{multirow}
\usepackage{stackengine}
\usepackage{hyperref}
\usepackage{array}
\usepackage{mathtools}
\usepackage{cases}
\usepackage{color}
\usepackage{array}
\usepackage{mathtools}
\usepackage{cases}
\usepackage{hyperref}
\begin{document}
\title{Interference Aware Path Planning for Mobile Robots in mmWave Multi Cell Networks}
\author{\parbox{6 in}{\centering Yijing Ren and Vasilis Friderikos\\
        Department of Engineering\\
        King's College London, London WC2R 2LS, U.K.\\
        E-mail:\{yijing.ren, vasilis.friderikos\}@kcl.ac.uk}
\vspace{-1.3\baselineskip}
}

\maketitle
\begin{abstract}
The emerging beyond 5G and envisioned 6G wireless networks are considered as key enablers in supporting a diversified set of applications for industrial mobile robots (MRs). The scenario under investigation in this paper relates to mobile robots that autonomously roam in an industrial floor and perform a variety of tasks at different locations whilst utilizing high directivity beamformers in mmWave small cells.
In such scenarios, 
the potential close proximity of mobile robots connected to different base stations, may cause excessive levels of interference having as a net result a decrease in the overall achievable data rate in the network. 
To resolve this issue, a novel mixed integer linear programming formulation is proposed where the trajectory of the mobile robots is considered jointly with the interference level at different beam sectors. 
Therefore,
creating a low interference path for each mobile robot in the industrial floor. 
A wide set of numerical investigations reveal that the proposed path planning optimization approach for the mmWave connected mobile robots can improve the overall achievable throughput by up to 31\%  
compared to an interference oblivious scheme, without penalizing the overall travelling time.
\end{abstract}
\IEEEpeerreviewmaketitle

\section{Introduction}
\IEEEPARstart{t}{he} recent advancements in wireless technologies and robotics have motivated the widespread application of networked mobile robots. Applications range from cargo transportation in smart factories \cite{factory} to
pick up and delivery in unmanned warehouses\cite{delivery}.
In these emerging highly dynamic industrial scenarios for manufacturing and industrial automation, widely known as Industry 4.0, 
the plethora of services that can be provided by mobile robots (MRs) require high data rates and reliability as well as low latency support\cite{industry,path}.
To this end, the use of a dense deployment of millimetre-wave (mmWave) small-cells can efficiently fulfil the above requirements for fully flexible production processes in smart factories using mobile robots\cite{IoT}. 
Moreover,
robots' path planning and orchestration \cite{yantong} are the key basis to accomplish a large set of tedious manufacturing processes that are often repetitive and as a result streamline workflows at the industrial floor. 
Therefore,
the optimization of the trajectory of the mobile robots should be considered jointly with the communication quality in order to improve the overall system performance.

The problem of MR path planning relates to the optimal robot movement between multiple nodes that involves target oriented decision-making.
Particularly,
in \cite{optimal},
a MR is required to visit all workstations/sensors to offload the requested data,
which is formulated as a Traveling Salesman Problem (TSP) to minimize the total travelling distance.
A mobile device scheduling approach is proposed in \cite{speed}
to minimize the data delivery latency with adjustable movement speed.
However,
only a single MR or device is considered in these prior works.
Additionally,
multiple MRs are used in \cite{method},
where routing strategy and path optimization are both studied to achieve multi-hop transmission capabilities.
Moreover, 
a trajectory planning model for multiple unmanned vehicles is designed in \cite{intell} to transport 
sensors to different locations while minimizing the total travel time.
Although the aforementioned studies strive to increase the overall network performance,
the strong interaction between path planning and achievable wireless throughput between MRs that might concurrently serve locations that create strong interference hasn't been explicitly investigated in the path planning optimization problem.

As already eluded above,
the network communication quality is another critical factor as it highlights the integration of path planning and wireless connection.
The authors in \cite{motion}
assign MRs to reach multiple locations and sustain communication with the base station (BS).
The trajectory planning problem for data-gathering MRs is presented in \cite{tour} to collect delay insensitive data with the minimal travel distance.
However,
the communication quality of service (QoS) cannot be guaranteed in \cite{motion,tour} to support a fully reliable transmission.
To this end,
in \cite{multi_mr},
a joint robots' path and robot-access point association planning is optimized to satisfy throughput requirements. 
A path planning problem for an aerial sensor network is studied in \cite{aerial} with connectivity constraints.

Considering the limitations in the aforementioned literature, in this work,
a novel MR path planning optimization scheme is proposed to explicitly avoid the downlink inter-beam interference (IBI) in mmWave wireless networks.
More specifically, the path planning of the different MR is performed in a way that strong IBI is eliminated in the network. Hereafter, when there is an IBI event we refer to that also as a communication collision.
The main contributions of this paper are as follows,
\begin{itemize}
\item An inter-beam interference free MR path planning optimization problem is proposed via a mixed integer linear programming (MILP) model;
\item A fully flexible Hamiltonian path is created for each MR, and time windows to visit each node location are also incorporated;
\item A wide set of numerical investigations reveal the significant gains of the proposed scheme compared with an inter-beam interference unaware path planning. 
\end{itemize}

The remainder of the paper is organized as follows. 
The system model is described in Section II.
In Section III, the path planning optimization problem for multiple MRs are formulated as a mixed integer linear program. Following that, section IV presents numerical investigations of the overall performance.
Finally, conclusions are drawn in Section V.

\section{System Model}
\begin{figure*}[htbp]
    \centering
    \subfigure[Pairwise interfering beams in mmWave cellular networks]{
    \label{subfig1}
    \includegraphics[width=0.4\textwidth]{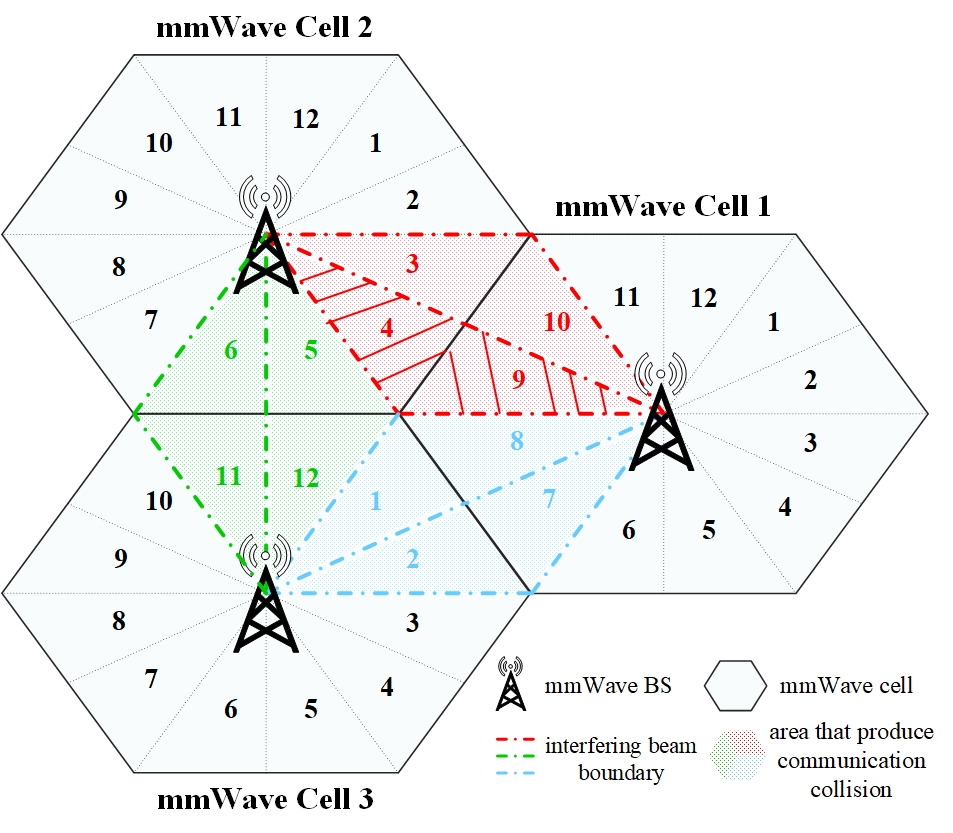}
    }
    \subfigure[The embedded MR trajectories]{
    \label{subfig2}
    \includegraphics[width=0.4\textwidth]{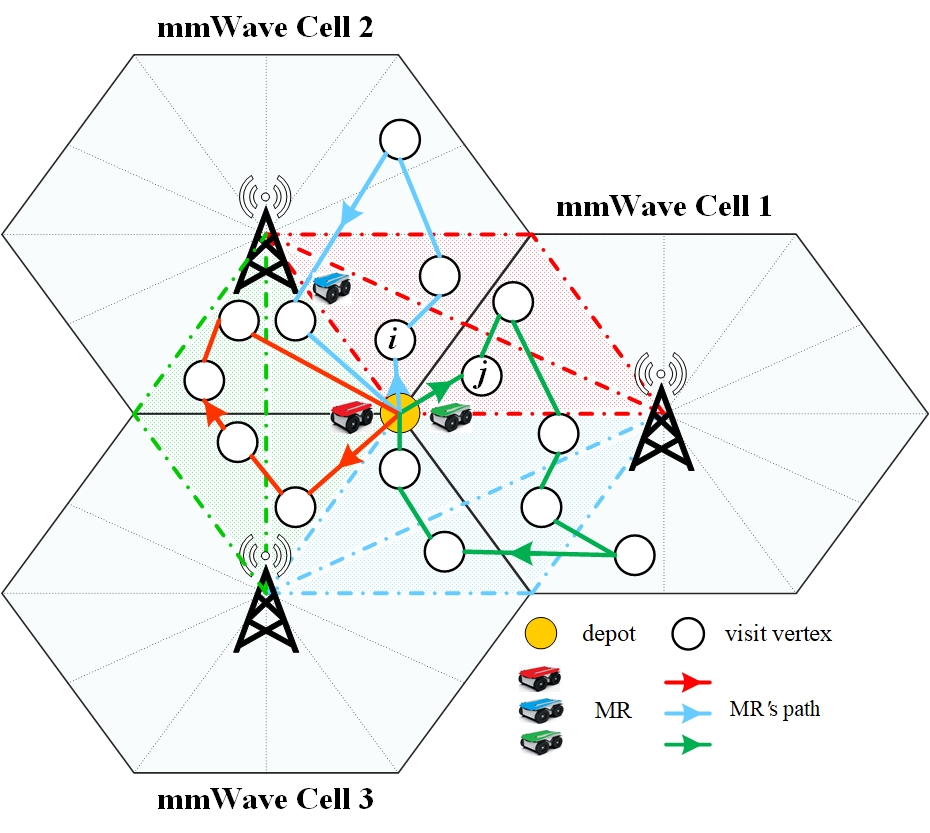}
    }
\caption{Illustration of the MR-assisted mmWave scenario}
\label{system model}
\end{figure*}
\subsection{MR-assisted MmWave Industrial Scenario}
The considered multi-cell mmWave communication industrial scenario is shown in Fig. \ref{system model}, 
where MRs are assigned to visit all different workstations to perform a specific task; hereafter those workstations are called as nodes.
Without loss of generality, the multi-cell mmWave network is constructed by identical hexagon cells equipped with beamformers, where BSs are deployed at the cell centers.
We denote the set of mmWave cells and BSs by $\mathcal{C}=\left\{1,2,\cdots ,C\right\}$ and $\mathcal{B}=\left\{1,2,\cdots ,B\right\}$, respectively.
The number of beams in every cell is denoted by $M$.
The set of visit nodes is randomly distributed within a pre-defined area, which relates to the industrial floor. 
We also assume that there is a centralized (edge) cloud center where all BSs are connected. 
In that respect, the exact location of all nodes is assumed to be known by all BSs.

As stated in \cite{graph},
a number of beams with certain directionality in a cell may cause inter-cell interference to a neighboring cell.
Following a similar assumption, we pair up the co-directional interfering beams between two adjacent cells,
such as $\left\{m_{2,4}, m_{1,9}\right\}$ as shown in Fig. \ref{subfig1}.
By the same token,
it is evident that there are 6 pairs of mutual interfering beams among the three cells,
and the colored hexagon can be recognized as an area where strong interference is taking place. 
We further define the simultaneous use of paired interfering beams in two neighbor cells as the \emph{communication collision} event,
which could cause severe interference and degrade the overall network performance.

To set the MRs' start and end points,
the junction of cells act as the depot.
As shown in Fig. \ref{subfig2},
we assume that all MRs dispatch from the depot and return to the (same) depot after visiting different nodes;
we further assume that each node is visited only once.
More specifically, 
each MR will serve the set of assigned nodes across the Hamiltonian path for a pre-defined service time utilizing a downlink mmWave transmission model.
When the MR has visited all the assigned nodes,
it returns to the depot for further instructions.


\subsection{Network Model}
The aforementioned mmWave wireless network with beamformers to support a set of mobile robots in an industrial setting can be modeled as an undirected graph $\mathcal{G}=(\mathcal{V},\mathcal{A})$\cite{undirected}.
The set of the MRs is defined by $\mathcal{K}=\left\{1,2,\cdots, K\right\}$.
$\mathcal{V}=\left\{0,1,\cdots ,v+1\right\}$ denotes the set of all nodes,
including the depot as MRs' start and end points.
The set of visiting nodes is denoted as $\mathcal{V}_p=\left\{1,2,\cdots ,v\right\}$.
$\mathcal{A}=\left\{(i,j): i,j \in \mathcal{V},i \neq v+1,j \neq 0,i \neq j\right\}$ is a set of links (routes) which
defines segments between different nodes.
We assume that MRs move with constant velocity $v$, and therefore
their travel time between node $i, j$ is denoted as $T_{ij}$,
which can be calculated by $T_{ij}=D_{ij}/v$.
Each node $i$ needs to be visited within a time horizon $[e_i,l_i]$.
When a MR reach node $i$, we assume that the service time, which is denoted as $w_i$, is pre-defined and constant.

As mentioned previously, if two MRs arrive simultaneously at nodes that covered by mutual interfering beams,
their immediate communication with BSs will cause communication collision.
Due to this,
we need to prevent MRs from visiting these nodes in the same time period.
Thus,
it is necessary to predetermine these visit node pairs.
A predefined matrix $h_{ij}$ denotes the spatial relation of visit nodes that can cause communication collision, i.e., strong inter-beam interference.
Specifically, 
$h_{ij}=1$ indicates that nodes $i$ and $j$ are individually located in the coverage of two mutual interfering beams,
meaning that these nodes should be avoided visited simultaneously,
otherwise $h_{ij}=0$.
Visiting node $i$ and $j$ can also be known as paired potential collision nodes.

\subsection{Wireless Downlink Transmission Model}
For the downlink transmission between mmWave BSs and MRs,
the antenna pattern proposed in \cite{manage,discrete} is adopted to model the beams in the BSs.
The directivity gain can be expressed as
\begin{equation}
\label{gain_beam}
G =
\begin{cases}
G_m = \frac{2\pi-(2\pi-\theta)G_s}{\theta},  & \text{in the main lobe},  \\
G_s,                                         & \text{in the side lobe},
\end{cases}
\end{equation} 
where $\theta$ is the operating beamwidth.
We assume that an omnidirectional antenna is used for MRs' reception \cite{UE} so $G^{Rx}=1$.
Moreover, the path loss models for both line-of-sight (LoS) and non-LoS (NLoS)
links are formulated as follows:
\begin{equation}
\label{path_loss}
L_s(D) =
\begin{cases}
\kappa_L D^{-{\alpha}_L},  & \text{with prob. } p_L(D)=e^{-\beta D},  \\
\kappa_N D^{-{\alpha}_N},  & \text{with prob. } p_N(D)=(1-p_L(D)),
\end{cases}
\end{equation}
where $s=\left\{\text{LoS, NLoS}\right\}$. 
$\kappa_L,\kappa_N$ are the path loss of LoS, NLoS links at the reference distance (1 meter) \cite{integrated,coverage_in}.
$\alpha_L$ and $\alpha_N$ are the pass loss exponents.
$p_L(D)$ is the LoS probability at distance $D$\cite{coverage_rate}.
Furthermore,
since the interference incurred by side lobes can be quite small,
we assume that the inter-beam interference is mainly caused by main lobes.
Therefore,
when MR $k$ is at visit node $i$, the output SINR is denoted as
\begin{equation}
\label{sinr}
S I N R_{i, k, b}=\frac{p_{m_c}G_{m_c}^{Tx}G^{Rx}\left\|h_{k,b}\right\|^{2}L_s(D_{i,k,b})}{\sum_{b^{\prime} \in \mathcal{B} \backslash \left\{b\right\}}^{B} I_{i,k,b^{\prime}}+N_0W}, 
\end{equation}
where $p_{m_c}$ is the transmission power allocated to beam $m_c$.
$\left\|h_{k,b}\right\|^{2}$ and $L_s(D_{i,k,b})$ denote the Nakagami fading \cite{coverage_rate} and path loss between BS 
$b$ and MR $k$. 
$N_0W$ and $W$ indicate the noise power and transmit bandwidth, respectively. 
Moreover,
the summation of $I_{i,k,b^{\prime}}$ denotes the interference power from all other BSs to MR $k$,
whereas in our case the strong interference only comes from one adjacent BS.
Thus, the achieved transmission rate between BS $b$ and MR $k$ at visit node $i$ is calculated by
\begin{equation}
\label{rate}
R_{i, k, b}=W\log_2(1+SINR_{i,k,b}). 
\end{equation}

\section{Problem Formulation}
In this section,
based on the defined scenario and system model,
we provide mixed integer linear programming (MILP) formulations for the following MR path planning problems.

\subsection{MR Path Planning Formulation}
A MILP formulation of the MR path planning problem with time windows is considered
in the mmWave cellular network,
which is called MR Path Planning with Communication Collision Unawareness (MP-CUA).
All MRs are deployed at the depot.
The decision variables of this model can be defined as follows:
\begin{equation}
\label{path}
x_{ijk} =
\begin{cases}
1,  & \text{if MR \emph{k} moves from node \emph{i} to node \emph{j}},  \\
0,  & \text{otherwise},
\end{cases}
\vspace{-0.1cm}
\end{equation} 
\begin{equation}
\label{time}
t_i = \text{the time which a MR arrives at node \emph{i}},
\vspace{-0.1cm}
\end{equation}
where $x_{ijk} \in \left\{0, 1\right\}$ and $t_i\geq 0$. 
The proposed MP-CUA can be seen as an extension of the vehicle routing problem with time windows (VRPTW)\cite{VRPTW}.
Hence, 
the constraints for creating MRs' Hamiltonian trajectories are enumerated as follows,
\begin{subequations}
\begin{equation}
\label{p1}
\begin{aligned} 
&\sum_{\substack{j \in \mathcal{V}_p}} x_{0jk}=1,&&\forall k\in \mathcal{K},\\ 
\end{aligned}
\vspace{-0.2cm}
\end{equation}
\begin{equation}
\label{p2}
\begin{aligned} 
&\sum_{\substack{j \in \mathcal{V}_p}} x_{jv+1k}=1,&&\forall k\in \mathcal{K},\\ 
\end{aligned} 
\vspace{-0.2cm}
\end{equation}
\begin{equation}
\label{p3}
\begin{aligned} 
&\sum_{\substack{k \in \mathcal{K}}} \sum_{\substack{i \in \mathcal{V}\\(i \neq j)}} x_{ijk}=1,&&\forall j\in \mathcal{V}_p,\\ 
\end{aligned}
\vspace{-0.2cm}
\end{equation}
\begin{equation}
\label{p4}
\begin{aligned} 
&\sum_{\substack{i \in \mathcal{V}\\(i \neq j)}} x_{ijk} - \sum_{\substack{h \in \mathcal{V}\\(h \neq j)}} x_{jhk}=0,&&\forall j\in \mathcal{V},\forall k\in \mathcal{K}.\\ 
\end{aligned}
\end{equation}
\end{subequations}
The constraints (\ref{p1}), (\ref{p2}) ensure that each MR must leave and return to the depot exactly once.
The constraint (\ref{p3}) guarantees that each node except for the depot is visited only once. 
The constraint (\ref{p4}) indicates that a MR arrives and leaves at a determined node.
Additionally,
the following constraints (\ref{tw1})-(\ref{tw4}) represent the required visiting time windows for MRs' path planning.
\begin{subequations}
\begin{equation}
\label{tw1}
\begin{aligned}
e_{0}+T_{0i}\leq t_{i}+M(1-x_{0ik}),&&\forall i\in \mathcal{V}_{p},\forall k\in \mathcal{K},\\
\end{aligned}
\end{equation}
\begin{equation}
\label{tw2}
\begin{aligned}
t_i+w_i+T_{ij}\leq t_{j}+M(1-x_{ijk}),&&\forall i,j(i \neq j)\in \mathcal{V}_{p},\\
&&\forall k\in \mathcal{K},\\
\end{aligned}
\vspace{-0.2cm}
\end{equation}
\begin{equation}
\label{tw3}
\begin{aligned}
t_i+w_i+T_{iv+1}\leq x_{iv+1k}(l_0-M)+M,&&\forall i \in \mathcal{V}_{p},\\
&&\forall k\in \mathcal{K},\\
\end{aligned}
\end{equation}
\begin{equation}
\label{tw_arriv1}
\begin{aligned}
t_i\leq e_{0}+T_{0i}+M(1-x_{0ik}),&&\forall i\in \mathcal{V}_{p},\forall k\in \mathcal{K},\\
\end{aligned}
\end{equation}
\begin{equation}
\label{tw_arriv2}
\begin{aligned}
t_j\leq t_i+w_i+T_{ij}+M(1-\sum_{\substack{k \in \mathcal{K}}} x_{ijk}),&&\forall i,j(i \neq j)\in \mathcal{V}_{p},\\
\end{aligned}
\vspace{-0.05cm}
\end{equation}
\begin{equation}
\label{tw4}
\begin{aligned}
e_i\leq t_i\leq l_i,&&\forall i \in \mathcal{V}_{p},
\end{aligned}
\end{equation}
\end{subequations}
where the big-M method is applied in these constraints and $M$ is a sufficiently large constant.
The constraint (\ref{tw1}) ensures the earliest time which MR $k$ arrives at node $i$.
The constraint (\ref{tw2}) ensures that MR $k$ arrives at a visit node $j$ after it has arrived and traveled from a visit node $i$.
The constraint in (\ref{tw3}) guarantees that a MR $k$ returns to the depot before the maximum window time $l_0$.
Note that the constraints in (\ref{tw1})-(\ref{tw3}) ensure no subtours of the designed paths\cite{subtour}.
The constraints in (\ref{tw_arriv1})-(\ref{tw_arriv2}) obtain the accurate arrival time of MRs at each visiting node.
The constraint (\ref{tw4}) ensures that node $i$ will be visited within the defined time window. 

Based on the above preliminaries, 
the MILP model for the MP-CUA problem that aims to minimize MRs' total travel time in an inter-beam interference free manner can be formulated as follows,
\begin{equation}
\begin{aligned} 
\label{CUA}
\text{MP-CUA:}~ \min_{\substack{\mathbf{X},\mathbf{t}}}\; \sum_{k \in \mathcal{K}}\sum_{i \in \mathcal{V}}\sum_{\substack{j \in \mathcal{V}\\(i \neq j)}} T_{ij}x_{ijk}
\end{aligned}
\end{equation}
\begin{subequations}
\begin{equation}
\label{cua1}
\begin{aligned} 
\text{s.t. (\ref{p1})-(\ref{p4}), (\ref{tw1})-(\ref{tw4})}, \\ 
\end{aligned}
\end{equation}
\begin{equation}
\label{x_binary}
\begin{aligned}
x_{ijk} \in \left\{0, 1\right\},&&\forall i,j \in \mathcal{V}, \forall k \in \mathcal{K},\\
\end{aligned}
\end{equation}
\begin{equation}
\label{t}
\begin{aligned}
t_i\geq 0,\quad \forall i \in \mathcal{V}_{p},\\
\end{aligned}
\end{equation}
\end{subequations}
where $\mathbf{X} \triangleq \left\{x_{ijk}|\forall i,j \in \mathcal{V},\forall k \in \mathcal{K}\right\}$ and $\mathbf{t} \triangleq \left\{t_i|\forall i \in \mathcal{V}_p\right\}$.
The constraint (\ref{x_binary}) and (\ref{t}) indicate the restrictive constraints for decision variable $x_{ijk}$ and $t_i$, respectively.  

\subsection{MR Path Planning Formulation with Communication Collision Awareness}
It is worth mentioning that the above optimization scheme is oblivious to the effects of communication collision,
which could happen if multiple MRs arrive and transmit in mutual interfering regions, i.e. paired potential collision nodes.
Consequently, these events would bring detrimental effects to the overall transmission data rate.
Therefore, 
MRs must visit paired potential collision nodes in a non-overlapping time manner in order to avoid communication collision.
To this end,
the formulation of communication collision aware path planning is proposed to improve the network communication performance.

As mentioned before,
$h_{ij}$ is a binary variable to denote whether $i,j$ are paired potential collision nodes.
Based on this,
we further define a variable $z_{ij}$ as follows,
\begin{equation}
\label{col}
z_{ij} =
\begin{cases}
1,  & \text{potential collision node \emph{i} is visited first}\\
    &\text{before \emph{j}},\\
0,  & \text{otherwise},
\end{cases}
\end{equation}
where $i,j(i \neq j) \in \mathcal{V}_c$ and $h_{ij}=1$.
$\mathcal{V}_c$ denotes the set of potential collision nodes which fall within the coverage of interfering beams.
To ensure that MRs visit pair-wise potential collision nodes in non-overlapping time windows,
the MR Path Planning with Communication Collision Awareness (MP-CA) formulation is given as follows,
\begin{equation}
\begin{aligned} 
\label{CA}
\text{MP-CA:}~ \min_{\substack{\mathbf{X},\mathbf{t}},\mathbf{Z}}\; \sum_{k \in \mathcal{K}}\sum_{i \in \mathcal{V}}\sum_{\substack{j \in \mathcal{V}\\(i \neq j)}} T_{ij}x_{ijk}
\end{aligned}
\vspace{-0.2cm}
\end{equation}
\begin{subequations}
\begin{equation}
\begin{aligned} 
\text{s.t.\quad (\ref{p1})-(\ref{p4}}), (\ref{tw1})-(\ref{tw4}), (\ref{x_binary})-(\ref{t})\\ 
\end{aligned}
\vspace{-0.2cm}
\end{equation}
\begin{equation}
\label{col1}
\begin{aligned} 
&\sum_{\substack{i \in \mathcal{V}_c}}\sum_{\substack{j \in \mathcal{V}_c}} z_{ij}=\frac{1}{2}\sum_{\substack{i \in \mathcal{V}_c}}\sum_{\substack{j \in \mathcal{V}_c}} h_{ij},
&&\forall i,j(i \neq j)\in \mathcal{V}_c,\\ 
\end{aligned} 
\vspace{-0.2cm}
\end{equation}
\begin{equation}
\label{col2}
\begin{aligned} 
&z_{ij}=1-z_{ji},&&\forall i,j(i \neq j)\in \mathcal{V}_c,\\ 
\end{aligned}
\vspace{-0.2cm}
\end{equation}
\begin{equation}
\label{col3}
\begin{aligned} 
&t_{i}+w_{i}\leq t_{j}+M(1-z_{ij})+M(1-h_{ij}),\forall i,j(i \neq j)\in \mathcal{V}_c,\\ 
\end{aligned}
\end{equation}
\begin{equation}
\label{col4}
\begin{aligned} 
&z_{ij} \in \left\{0, 1\right\},&&\forall i,j(i \neq j)\in \mathcal{V}_c,\\ 
\end{aligned}
\vspace{-0.2cm}
\end{equation}
\begin{equation}
\label{h}
\begin{aligned} 
&h_{ij} \in \left\{0, 1\right\},&&\forall i,j(i \neq j)\in \mathcal{V}_c,\\ 
\end{aligned}
\end{equation}
\end{subequations}
where $\mathbf{Z} \triangleq \left\{z_{ij}|\forall i,j(i \neq j)\in \mathcal{V}_c\right\}$.
The constraint (\ref{col1}) defines the total number of collision node pairs.
The constraint (\ref{col2}) ensures that either potential collision node $i$ is visited before $j$ or by contrary.
Constraint (\ref{col3}) guarantees the visiting time windows of both potential collision nodes do not overlap if $z_{ij}$ and $h_{ij}$ are 1.
The constraint (\ref{col4}) and (\ref{h}) indicate that $z_{ij}$ and $h_{ij}$ are both binary variables.

Overall, 
the MP-CUA and MP-CA problems are in the form of a MILP,
and each one of them is a generalization of both the TSP and the Vehicle Routing Problem (VRP).
Moreover, 
these problems optimize a TSP-like trajectory with the minimal total travel time.
Therefore,
both optimization problems resemble a TSP or VRP in problem formulation and computational complexity,
meaning that they are NP-hard problems\cite{NP}.

\section{Numerical Investigations}
In this section,
numerical investigations are presented to evaluate the performance of the proposed schemes.
Without loss of generality, 
three hexagon cells are considered and the total number of beams per cell is $M=12$.
The side length of each cell is set as $l=50\;m$ and the number of MRs is 3.
All other simulation parameters are summarized in Table \ref{parameter}.
\begin{table}[ht]
\scriptsize
\caption{\label{tableI} Simulation Parameters.} 
\centering
\begin{tabular}{|l|c|}
\hline
\textbf{Parameter}&\textbf{Value} \\
\hline
\hline
Bandwidth ($W$) & 100 MHz\\
\hline
Transmission power per mmWave BS ($p_c$) & 20 dBm\\
\hline
Directivity gain in main lobes ($G_{m}$) & 10.9\\
\hline
Directivity gain in side lobes ($G_{s}$) & 0.1\\
\hline
Path loss exponents of LoS, NLoS links ($\alpha_L$,$\alpha_N$) & $\alpha_L$=2 $\alpha_N$=4\\
\hline
Path loss of LoS, NLoS links at a distance of 1 meter ($\kappa_L$,$\kappa_N$) & $(\frac{c}{4\pi f_c})^2$ \\ 
\hline
Noise power spectral density ($N_0$) & -174 dBm/Hz\\
\hline
Carrier frequency ($f_c$) & 28 GHz\\
\hline
Beamwidth ($\phi_b$) & $\frac{\pi}{6}$\\
\hline
Velocity of MRs ($v$) & 5 m/s\\
\hline
LoS probability factor (1/$\beta$) & 141.4 m\\ 
\hline
\end{tabular}
\label{parameter}
\end{table}

\begin{figure}[ht]
\centering
\subfigure[Paths of MRs under MP-CUA]{
\label{path_1}
\includegraphics[width=0.43\textwidth]{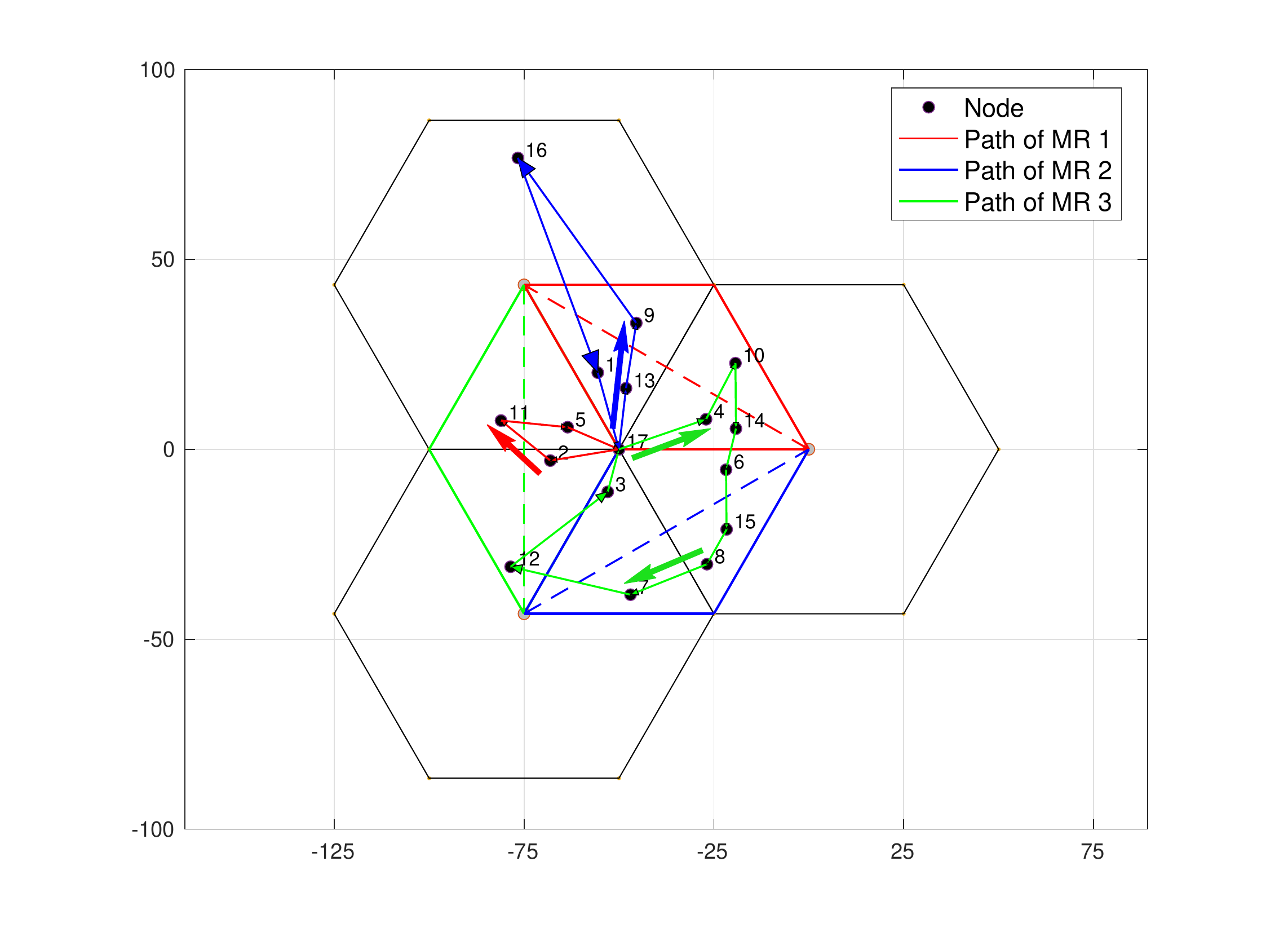}}
\subfigure[Paths of MRs under MP-CA]{
\label{path_2}
\includegraphics[width=0.43\textwidth]{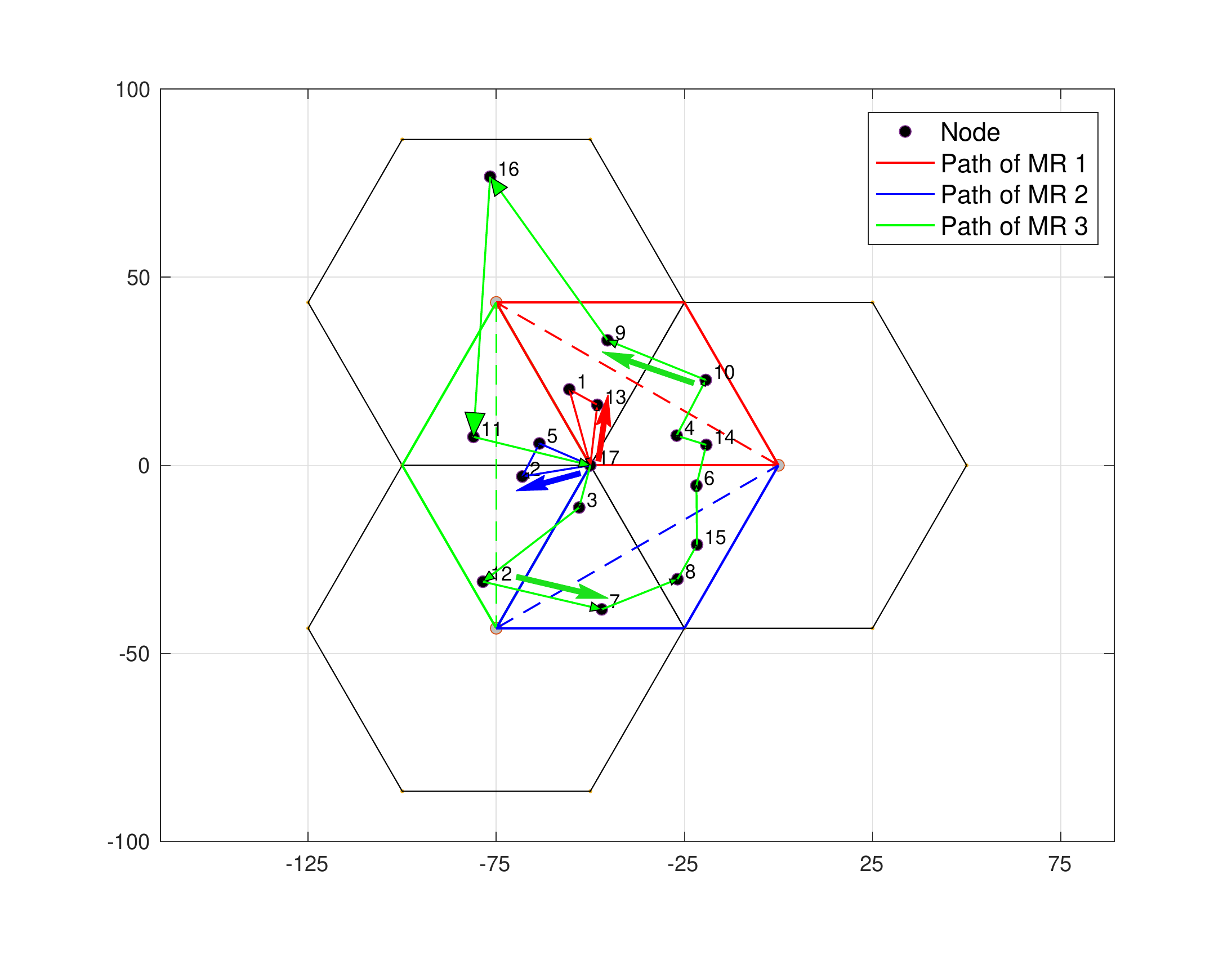}}
\caption{Path planning for 3 MRs, 16 visit nodes}
\label{MPP}
\end{figure}

In order to demonstrate how the MP-CA scheme is able to avoid communication collisions,
a detailed example of 3 MRs and 16 visiting nodes is first presented before elaborating on Monte Carlo simulations.
As shown in Fig. \ref{path_1},
because MRs visit $\left\{9,10\right\}$ and $\left\{4,13\right\}$ in overlapping time windows,
the communication collision occurs respectively at both locations during robots' movement.
Afterwards,
in Fig. \ref{path_2},
all MRs have altered their paths while ensuring no communication collision happened,
which brings an increase in the overall transmission rate.
Due to the changing paths,
the total travel time has increased by 0.508\;s.
Through observing various simulation cases,
although the change in trajectory direction can achieve collision avoidance,
it is more likely that MRs' paths diverge from the shortest ones in order to avoid communication collision.

As mentioned before,
the simultaneous transmission between two MRs to connected BSs could cause communication collision if they lie within the coverage of mutual interfering beams.
Based on this,
we can define the colored hexagon as a collision area that contains the interfering beams of adjacent cells.
\begin{figure}[htbp]
\centering
\subfigure[Scenario A]{
\label{Scenario A}
\includegraphics[width=4cm]{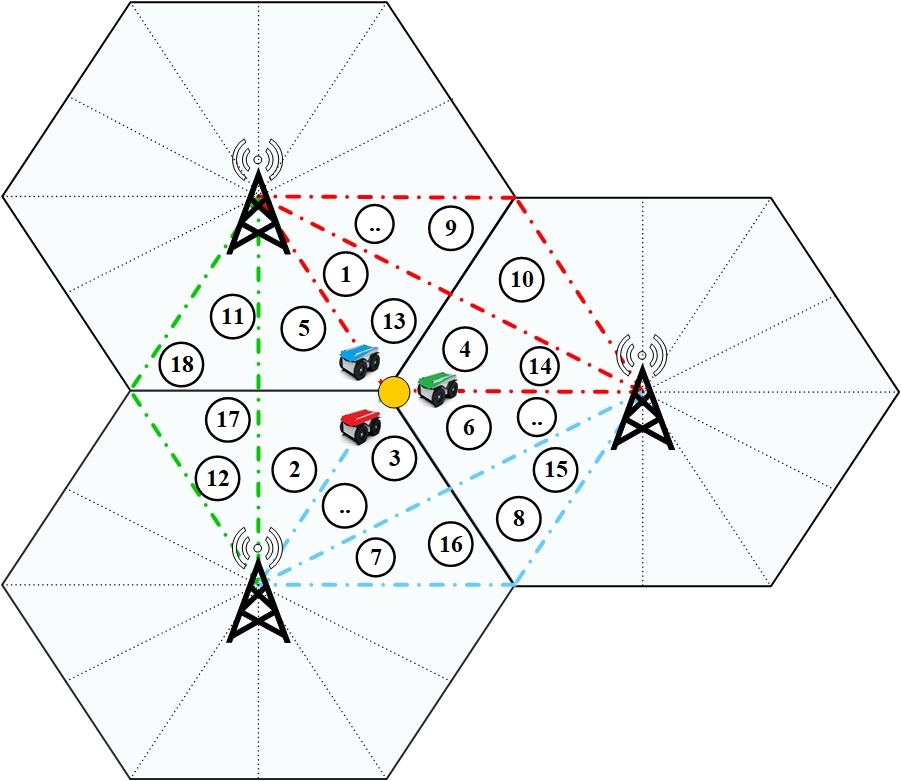}}
\subfigure[Scenario B]{
\label{Scenario B}
\includegraphics[width=4cm]{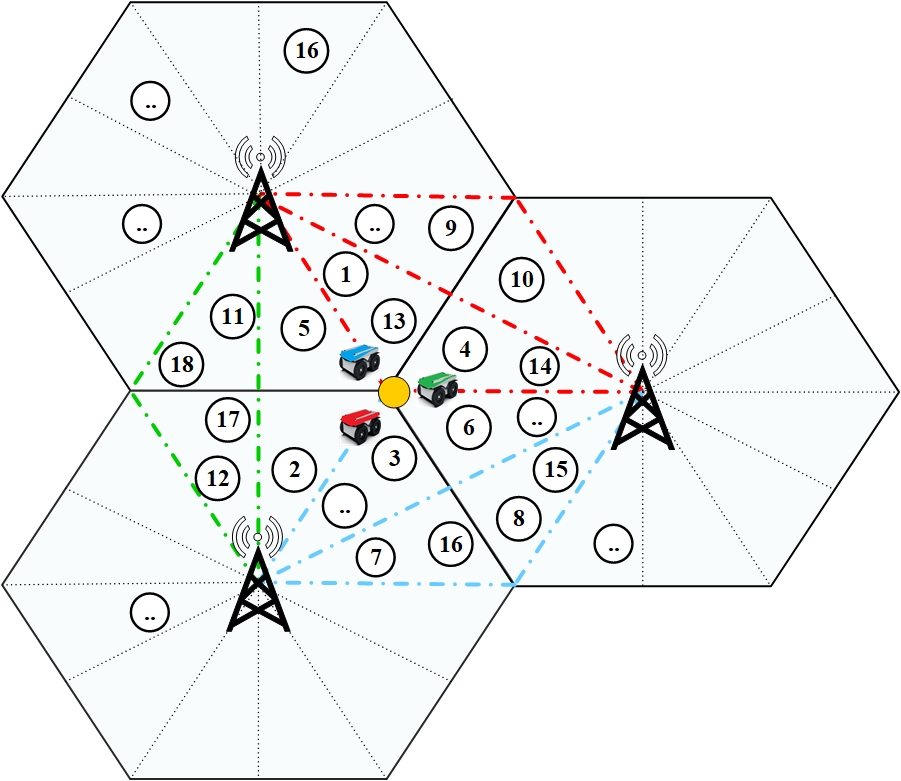}}
\caption{Node Distribution Setting}
\label{Node distribution scenario}
\vspace{-3mm}
\end{figure}
Two spatially distribution scenarios of nodes that are required to be visited by the MRs are depicted in Fig. \ref{Node distribution scenario}.
In scenario A, all nodes to be visited are distributed in the collision area. 
In scenario B, the nodes are scattered uniformly among the coverage area of the three cells.
All reported numerical results are obtained by averaging 100 Monte Carlo simulations.

\begin{figure}[ht]
    \centering
    \includegraphics[width=0.43\textwidth]{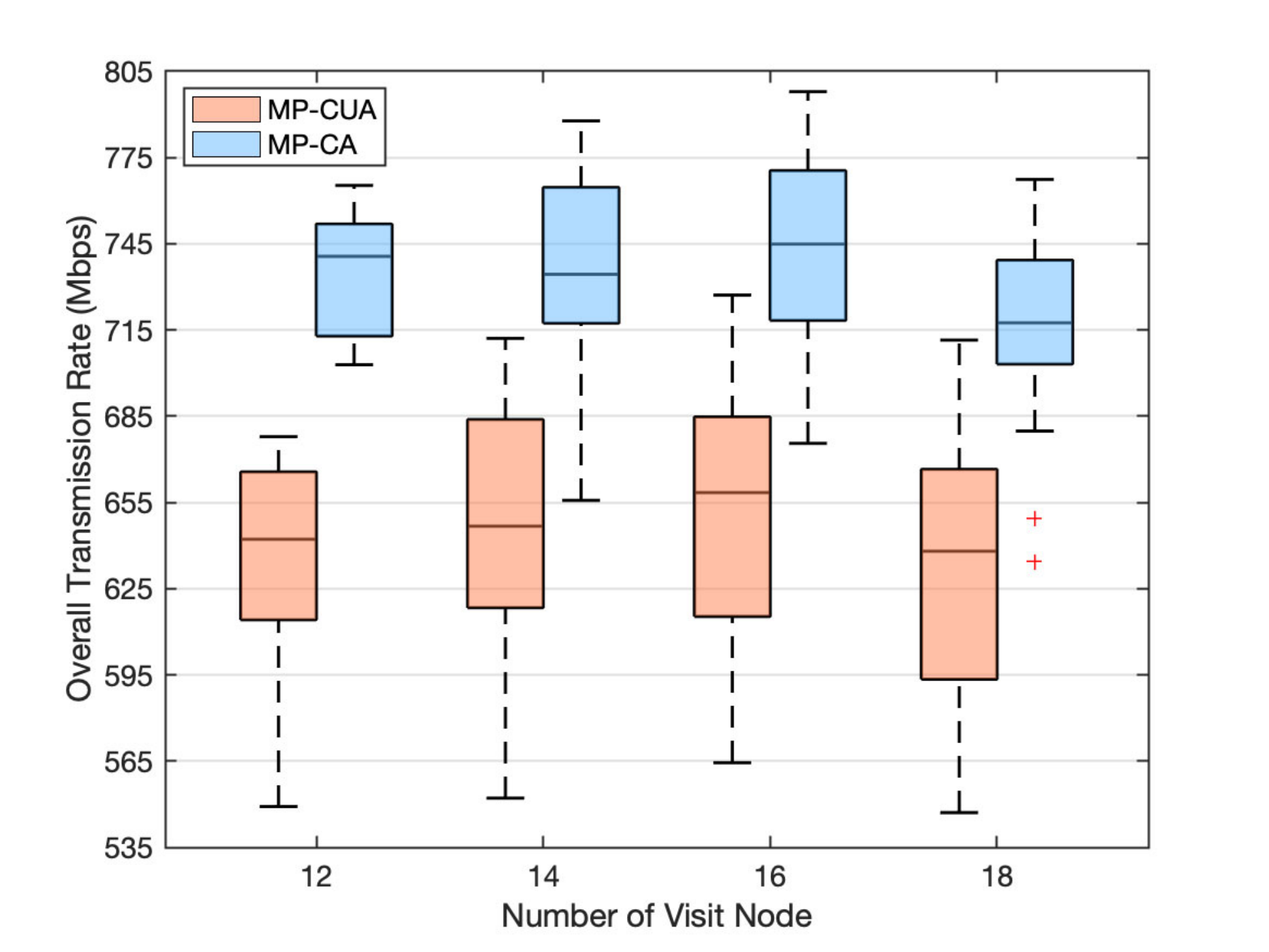}
    \caption{Overall transmission rate comparison in Scenario A}
    \label{rate_1}
\end{figure}

To evaluate the impact of communication collision,
we first focus on the performance in scenario A.
With respect to different number of visiting nodes,
Fig. \ref{rate_1} presents the overall transmission rate distribution for the MP-CUA and MP-CA schemes. 
The overall transmission rate is calculated as the average achievable data rate at each visiting node.
We note that in Fig. \ref{rate_1},
for each box-whisker plot,
the top and bottom lines of whiskers represent the maximum and minimum rate, respectively.
It can be observed that the achievable transmission rate of the MP-CUA scheme range
between 595 to 685 Mbps,
and the MP-CA scheme gains a better performance than MP-CUA in the order of 95 Mbps.
Moreover,
compared with the MP-CUA for the cases of 12, 14, 16, 18 visiting nodes,
the MP-CA scheme can improve the overall transmission rate
by an average of $15.86\%$, $13.73\%$, $14.49\%$, and $13.71\%$, respectively. 
The maximum gain for the overall transmission rate mounts up to $31.93\%$ when there are 16 visiting nodes. Also, compared to the MP-CUA,
the MP-CA scheme achieves at least an increase of 50.4 Mbps with a corresponding gain of $7.34\%$.
The reason behind these gains is due to the elimination of communication collisions,
which is ignored under the MP-CUA scheme since it is oblivious of the effects caused by the simultaneous use of mutual interfering beams.
Therefore,
the occurrence of such communication collisions result in a significant decrease in the overall data transmission rates,
while showing the effectiveness of the proposed MP-CA method. 
In addition,
the total travel time comparison is shown in Table \ref{time_1}.
Observe that the time consumed by both schemes are generally similar.
\vspace{-0.2cm}
\begin{table}[htbp]
\centering
\caption{Total travel time comparison in Scenario A}
\label{time_1}
\begin{tabular}{|c|c|c|c|c|}
\cline{1-5}
& n\_visit=12 & n\_visit=14 & n\_visit=16 & n\_visit=18\\
\hline
MP-CUA & 61.4661s & 63.0041s & 65.3630s & 68.4273s\\
\hline
MP-CA & 61.4661s & 63.0201s & 65.3936s & 68.4534s\\
\hline
\end{tabular}
\end{table}

Furthermore,
the overall achievable transmission rates and travel times in scenario B is investigated as well. 
Unlike scenario A,
some nodes are allocated in the non-collision area,
where the achievable data rate is not affected by inter-beam interference, i.e., what we call communication collision.
\begin{figure}[htbp]
    \centering
    \includegraphics[width=0.42\textwidth]{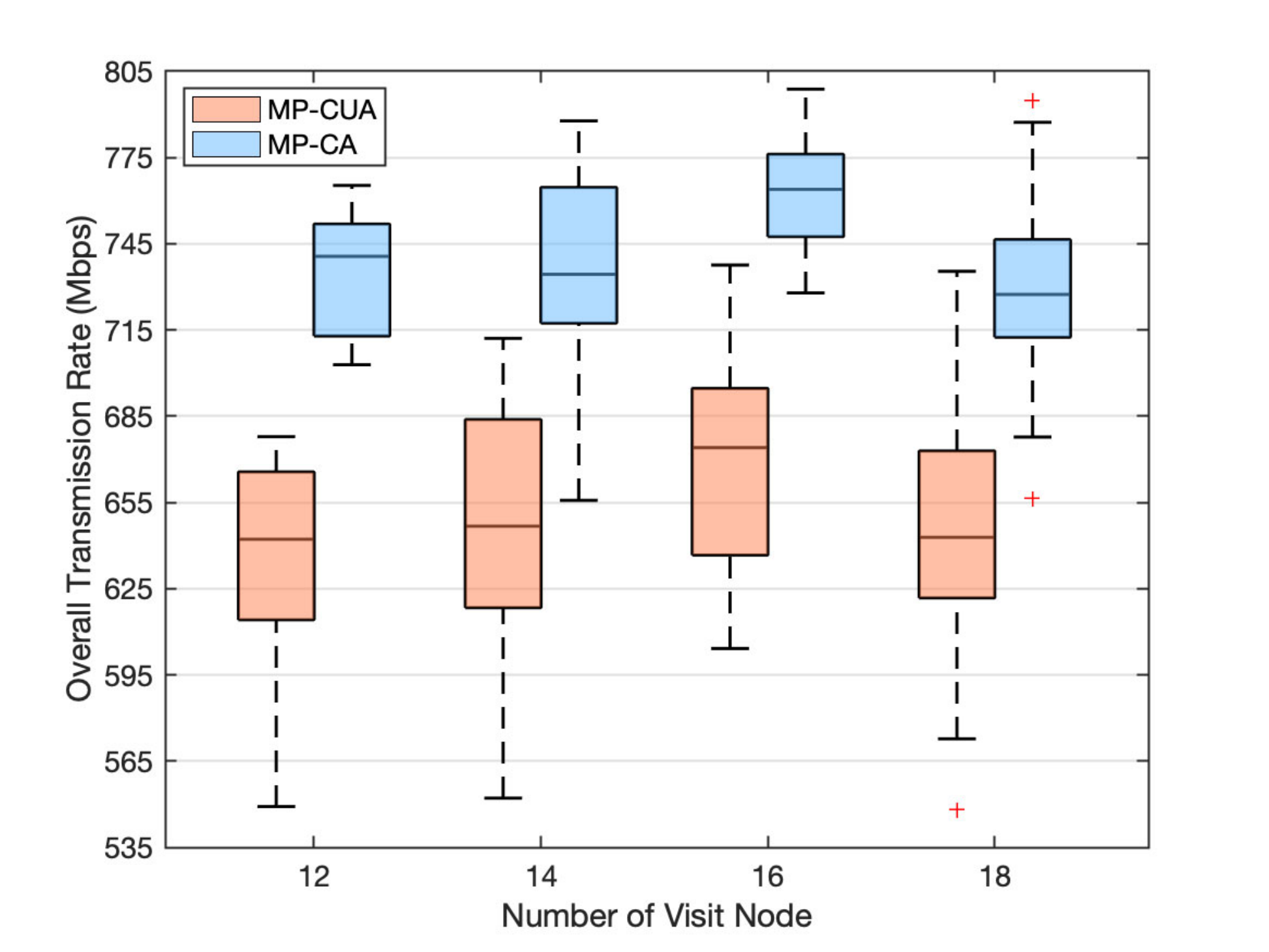}
    \caption{Overall transmission rate comparison in Scenario B}
    \label{rate_2}
\end{figure}
As shown in Fig. \ref{rate_2},
compared to the MP-CUA scheme for different number of visiting nodes (12, 14, 16 and 18),
the proposed  MP-CA scheme can improve the data rate performance by $15.86\%$, $13.73\%$, $15.26\%$ and $13.79\%$, respectively.
\begin{table}[htbp]
\centering
\caption{Total travel time comparison in Scenario B}
\label{time_2}
\begin{tabular}{|c|c|c|c|c|}
\cline{1-5}
& n\_visit=12 & n\_visit=14 & n\_visit=16 & n\_visit=18\\
\hline
MP-CUA & 61.4661s & 63.0041s & 76.5636s & 79.1342s\\
\hline
MP-CA & 61.4661s & 63.0201s & 76.6237s & 79.1944s\\
\hline
\end{tabular}
\end{table}
The reason behind the fluctuating gains is that with the increase number of nodes, 
the probability that a number of nodes are in the non-collision area is higher. 
However, the maximum gain of the overall achievable transmission rate is $30.46\%$.
Additionally,
in Table \ref{time_1} and \ref{time_2},
it can be revealed that the time gap between MP-CUA and MP-CA is reflected in the decimal part.
This is because nodes are densely distributed within a hexagon area,
indicating that the distance difference between nodes can be deemed as small.
Therefore,
it results in minor time difference between the two schemes.


\section{Conclusions}
Mobile robots (MRs) that can roam autonomously in an industrial floor and perform a variety of tasks are expected to revolutionize industrial automation owing to the benefits that they can bring, such as high efficiency and safety for humans. 
This paper studies a novel problem of MR Hamiltonian path planning with communication collision awareness (MP-CA),
which is formulated as a MILP model with time windows.
Compared to MR path planning with communication collision unawareness (MP-CUA),
MP-CA considers in an explicit manner the inter-beam interference in a mmWave network and provides trajectories for the MRs that avoid communication collision.
A wide set of numerical investigations reveal that the proposed MP-CA scheme achieves an average and a maximum of more than 15\% and 31\% higher transmission rate respectively, while requiring similar travel time compared to a trajectory that does not consider mmWave interference.

\bibliographystyle{IEEEtran}
\bibliography{IEEEabrv,reference} 
\end{document}